\documentclass[reprint,amsmath,amssymb,aps]{revtex4-1}
\usepackage[utf8]{inputenc}
\usepackage{physics}
\usepackage{graphicx}
\usepackage{dcolumn}
\usepackage{bm}
\usepackage{amssymb}
\usepackage{braket}
\usepackage{mathtools}
\usepackage{comment}
\usepackage{epstopdf}

\usepackage{amsmath}
\usepackage{tikz}

\begin{document}

\graphicspath{ {./images/} } 
\title{The dependence of evanescent wave polarization on the losses of guided optical modes}
\author{Sinuh\'e Perea-Puente}
\author{Francisco J. Rodr\'iguez-Fortu\~no}%
 \email{Corresponding author: francisco.rodriguez-fortuno@kcl.ac.uk.}

\affiliation{
Department of Physics, King's College London, Strand, London WC2R 2LS, United Kingdom\\}

\date{\today}

\begin{abstract}
Spin-momentum locking of evanescent waves describes the relationship between the propagation constant of an evanescent mode and the polarization of its electromagnetic field, giving rise to applications in light nano-routing and polarimetry among many others. The use of complex numbers in physics is a powerful representation in areas such as quantum mechanics or electromagnetism; it is well known that a lossy waveguide can be modeled with the addition of an imaginary part to the propagation constant. Here we explore how these losses are entangled with the polarization of the associated evanescent tails for the waveguide, revealing a well-defined mapping between waveguide losses and the Poincar\'e  sphere of polarizations, in what could be understood as a ``polarization-loss locking'' of evanescent waves. We analyze the implications for near-field directional coupling of sources to waveguides, as optimized dipoles must take into account the losses for a perfectly unidirectional excitation. We also reveal the potential advantage of calculating the angular spectrum of a source defined in a complex, rather than the traditionally purely real, transverse wavevector space formalism.
\end{abstract}

\maketitle

\section{Introduction}

Since the birth of {nanoscience} in the latter decades of the twentieth century, it is possible to revisit some old well-established theoretical concepts and exploit them for novel {near-field} applications in subwavelength phenomena. In particular, consider the case of  evanescent waves \cite{evanescentmedic}, known more than 150 years ago and traditionally considered as a mere theoretical corollary in total internal reflection situations. In recent years, evanescent waves have become deeply re-envisioned as fascinating and promising tools for optical applications in the nanoscale. Apart from carrying linear and angular momentum in the direction of propagation, evanescent waves were lately shown to also transport a transverse spin angular momentum \cite{originspin,aiello,basicspin, wheels}, leading to spin-momentum locking \cite{Bliokh2015a,evanescent} (also known as photonic quantum spin hall effect \cite{Bliokh2015a} in line with its electronic counterpart). 

Spin-momentum locking is one of the most promising features of evanescent waves, as it is an inherent property independent of their source. It has been experimentally demonstrated that spin-momentum locking occurs in a wide range of physical systems such as  surface-plasmon-polaritons \cite{pakscience}, optical fibers \cite{evanescent,fib1,fib2} and silicon waveguides \cite{copy1}. It plays a key role enabling selective coupling in the near and far field via polarized dipoles \cite{pakscience,angularmomen}, giving rise to recoil {optical forces} \cite{engheta,opticalforce,jackforce,recoil} and presents further applications in other areas such as optical isolation \cite{iso1,iso2}, nanopolarimetry \cite{nanopo}, or optical vortex emitters \cite{vorte}. 
We also note that the underlying physics applies to wave-fields beyond electromagnetism such as acoustics \cite{Wei2020,accoustics} and gravitational waves \cite{gravitation}.

\begin{figure}[!htbp]
    \includegraphics[width=8cm]{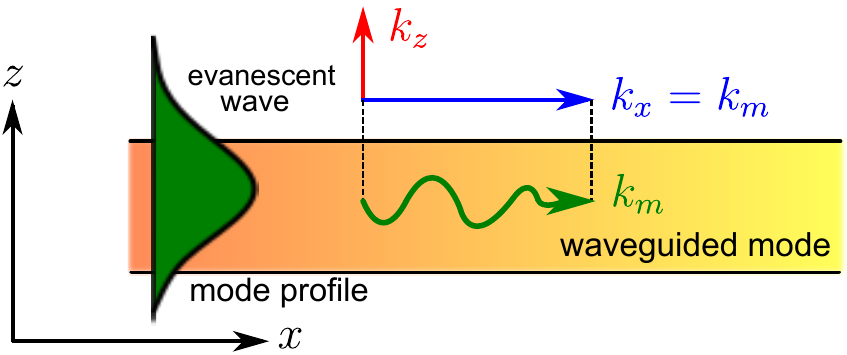}
    \caption{Evanescent tail of a 2D-slab waveguide showing the relation between the evanescent wavevector $\mathbf{k}=k_x \mathbf{\hat{x}} + k_z \mathbf{\hat{z}}$ and the waveguide propagation constant $k_m$.}
     \label{fig:concept}
\end{figure}

The spin-momentum locking of evanescent tails in a waveguided mode ultimately stems from the transversality condition of momentum eigenmodes, $\mathbf{k}\cdot\mathbf{E}=\mathbf{k}\cdot\mathbf{H}=0$, relating the wavevector $\mathbf{k}$ to the electric field $\mathbf{E}$ and magnetic field $\mathbf{H}$ polarization of the mode \cite{Bliokh2015a}. Therefore, the propagation constant of the waveguide mode $k_m$ is crucial, because momentum conservation in translationally-invariant waveguides requires that the component of the evanescent field's wavevector in the propagation direction, $k_x$, must be equal to the intrinsic propagation constant of the mode, as depicted in Fig. \ref{fig:concept}. In recent literature \cite{aiello,evanescent}, lossless waveguide modes are typically considered - which means that the propagation constant $k_m$, and hence $k_x$ too, is taken as a real number. In this well-known situation, the total wavevector of the evanescent wave still exhibits complex-number behavior due to the wavevector $\mathbf{k}$ having an imaginary component in the perpendicular direction to the guided mode $k_z$, corresponding to the direction of evanescent attenuation, while having a purely real component in the propagation direction, $k_x$. As we know, both components are related via the wave-equation $\mathbf{k}\cdot\mathbf{k} = k^2$, where $k = n\omega/c$ is the background wavenumber for a medium with refractive index $n$. However, more degrees of freedom can be gained if one considers complex propagation constants corresponding to lossy waveguides. Mathematically, a lossy waveguide is simply associated with a complex propagation constant. This, in turn, implies a complex wavevector component in the propagation direction $k_x = k_x' + i k_x''$ for the evanescent wave. In this case, to satisfy the wave-equation, both $k_x$ and $k_z$ wavevector components acquire both real and imaginary parts, which therefore affects the polarization properties and the spin of the associated evanescent waves via the transversality conditions. While this is an expected result, or at least should not be surprising, in this work we wish to study the phenomenon in depth, to uncover its subtleties. In particular, we will see that the presence of losses must be taken into account when designing a dipole for optimal directionality in evanescent coupling.

This work is split into two main parts. First, we will analyze the polarization in the evanescent tail for a lossy waveguide. We will study the geometric paths described in the Poincar\'e sphere by this polarization as the losses of the waveguide are varied. Secondly, we will study evanescent coupling between a dipole source and a lossy waveguide, exploring the effects of loss in the guided mode, and how the dipole optimization and tunability for selective control of unidirectional excitation should be re-calculated in this scenario. For this, we will use both Fermi's golden rule \cite{fermi,fermi2} and the angular spectrum approach \cite{pakscience,angularmomen,jackforce,Wei2020,multipole}, which we will here extend to a complex domain. We will prove that the dipole polarization must be re-optimized taking into account the losses for a perfect contrast directionality and this optimization can be associated with a zero in a complex domain of the angular spectrum.

\section{Polarization paths for the evanescent tails of a lossy waveguide}

In our first approach, we will calculate the polarization ellipse for the evanescent tails of a lossy waveguide. In order to get at the essence of the phenomenon, we will study the simplest possible scenario, a two-dimensional problem as shown in Fig. \ref{fig:concept}, where a slab waveguide is embedded in an infinite-homogeneous background of refractive index $n$. In our calculations we take $n=1$ for simplicity, i.e., free space surroundings. The lossy waveguide supports a well defined time-harmonic mode defined by its propagation constant  $k_m = k_m' + i k_m''$. The propagator in the waveguide is given by $e^{i(k_m x - \omega t)} = e^{i k_m' x}e^{-i \omega t} e^{- k_m'' x}$, clearly exhibiting phase propagation in space associated to the propagation constant $k_m'$, time-harmonic phase advance in time due to the real-valued angular frequency $\omega$, and an evanescent amplitude decay in space corresponding to the attenuation constant $k_m''$ caused by waveguide mode losses. The evanescent tails of such a mode can be written as a momentum eigenvector in complex phasor notation as $ \{ \mathbf{E}_\mathrm{ev}(\mathbf{r}), \mathbf{H}_\mathrm{ev}(\mathbf{r}) \}  = \{ \mathbf{E}_0, \mathbf{H}_0 \} e^{i \mathbf{k} \cdot \mathbf{r}}$, where $\mathbf{E}_0$ and $\mathbf{H}_0$ are the evanescent wave electric and magnetic field polarization, $\mathbf{k}$ is the wavevector of the evanescent wave, and $\mathbf{r}$ is the position vector. Such a momentum eigenvector must be a solution to Maxwell's equations, and as such it must fulfill two important requirements:

\begin{equation} \label{eq:generalreq}
\mathbf{k} \cdot \mathbf{k}=k^2 \quad \mathrm{and} \quad
\mathbf{k} \cdot \mathbf{E}_\mathrm{ev} = \mathbf{k} \cdot \mathbf{H}_\mathrm{ev} = 0.
\end{equation}

The first requirement comes from the homogeneous Helmholtz wave-equation derived \cite{max} from Maxwell's equations, and the second comes from Gauss' law in the absence of sources, also known as the transversality condition \cite{evanescent}. Note, as is well known, that the first equation acts on a complex wavevector, so it is not the analytical equation of a circumference.

To simplify the situation further we will consider only a transverse-magnetic (TM or p-) mode, in which the magnetic field polarization of the evanescent wave is trivial $\mathbf{H}_0=H_y \mathbf{\hat{y}}$ and the electric field is responsible for all the interesting polarization phenomena and transverse spin $\mathbf{E}_0 = E_x \mathbf{\hat{x}} + E_z \mathbf{\hat{z}}$. This apparent loss of generality is justified because a transverse-electric (TE or s-) mode would show identical phenomena, but simply switching the roles between $\mathbf{E}_0$ and $\mathbf{H}_0$. Hence, in our simplified case of a TM mode and 2D problem $\mathbf{k}=k_x \mathbf{\hat{x}} + k_z \mathbf{\hat{z}}$, the above conditions in Eq. \ref{eq:generalreq} can be simplified to:

\begin{equation} \label{eq:simplifiedreq}
k_x^2 + k_z^2 = k^2 \quad \mathrm{and} \quad
k_x E_x + k_z E_z = 0.
\end{equation}

\begin{figure*}[!htbp]
    \includegraphics[width=1\textwidth]{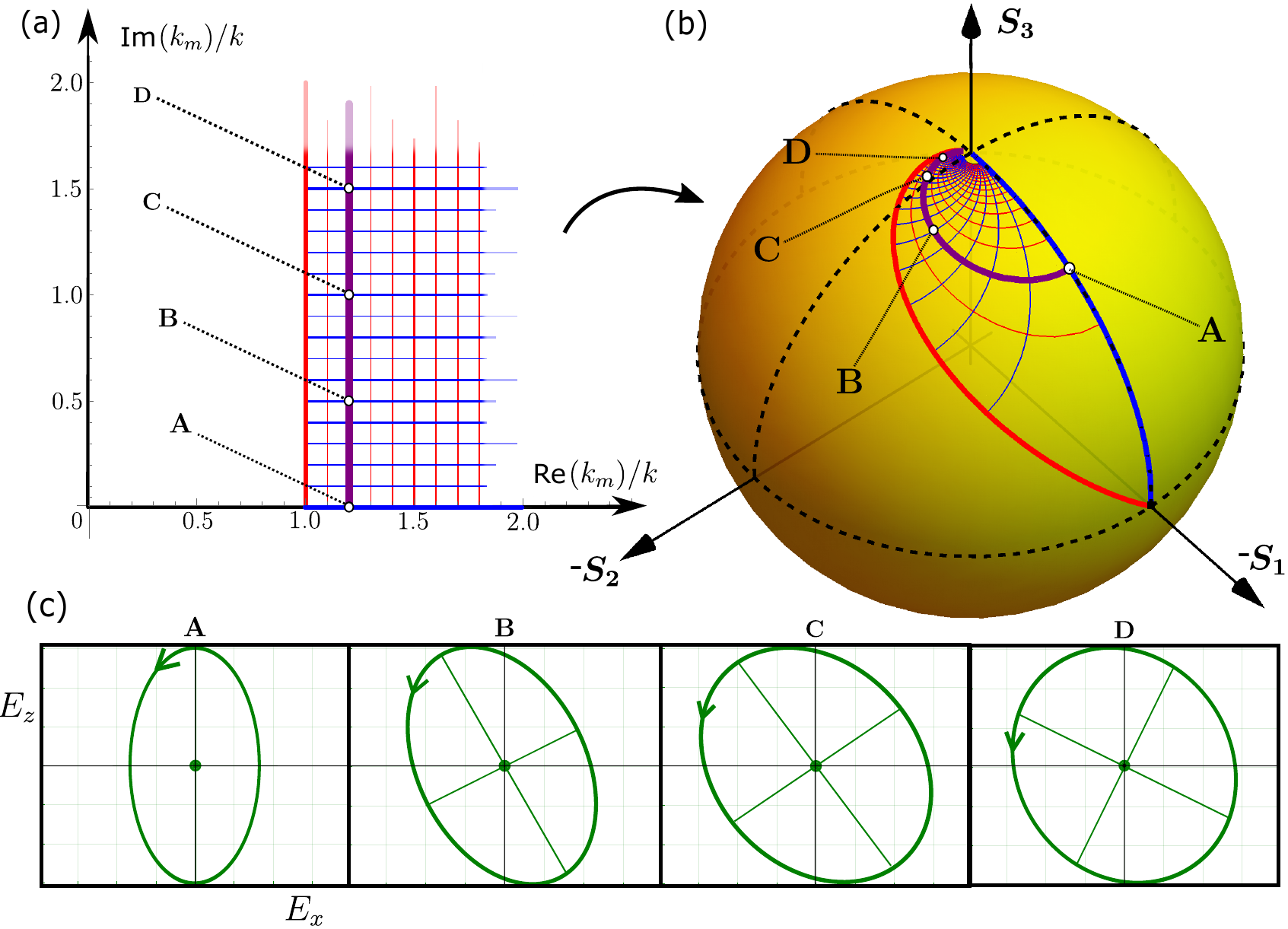}
    \caption{Evanescent wave polarization in Poincar\'e sphere from normalized Stokes parameters $\{S_1,S_2,S_3\}/S_0$ (b) as the propagation constant of the waveguide mode is varied in the complex plane (a) where the red and blue lines represent a mesh in the complex $k_m$ plane, mapped into the Poincar\'e sphere. The thick blue line represents the line $k_m \in [1,\infty)$ while the red and purple thick lines represent varying imaginary parts for $k'_m/k=1$ and $k'_m/k = 1.2$ respectively. A selection of four points $\{A,B,C,D\}$ have been chosen as an example, with propagation constants $k_m/k = 1.2 + \{0,1,2,3\}\frac{i}{2}$ showing the influence of the imaginary part of $k_m$ (waveguide losses, $k''_m$) on the polarization. The corresponding electric field polarization ellipse is shown in (c).}
\label{fig:poincare}
\end{figure*}

\noindent With these equations, together with the fact that $k_x = k_m$ due to conservation of momentum parallel to the axes of translational invariance in the waveguide, we are able to study all the changes in the polarization of the evanescent tails with the addition of losses to the waveguide. In order to illustrate this behavior, we map a grid in the complex plane of propagation constants (Fig. \ref{fig:poincare}(a), corresponding to any possible propagating mode) to the associated transverse polarization that the evanescent tail would have for that mode, in the Poincar\'e sphere, as shown in Fig. \ref{fig:poincare}(b) (see Appendix A for detailed calculations). The plane of polarization ellipses used to calculate the Stokes parameters to depict the Poincar\'e sphere is taken as the $(x,z)$ plane, parallel to the propagation direction $x$, as expected for the electric field of a p-polarized evanescent mode.

To analyze Fig. \ref{fig:poincare}, we highlight as a blue line the well-known case of lossless waveguided modes, corresponding to a real $k_m/k \in [1,\infty)$. When $k_m/k=1$, this is not a guided mode but a propagating plane wave, with linear p-polarization, hence we are in the equator of the Poincar\'e sphere. When $k_m/k>1$ is increased, making the wave more and more evanescent (i.e. decaying more strongly), the polarization of the evanescent wave follows a geodesic path, moving towards the upper pole, where $S_3/S_0 = 1$ corresponding to purely circular polarization in the Poincar\'e sphere, in agreement with the well-known appearance of a transverse spin. When losses are added, it is interesting that the polarization moves away from the $S_2 = 0$ condition \cite{evanescent}, as $k_z$ is not purely imaginary, characteristic of lossless modes. The presence of non-zero $S_2$ indicates a tilting of the polarization ellipse due to the losses. As losses are increased, the polarization follows a cardioid-like path in the Poincar\'e sphere, also tending to the upper pole in the limit of high losses.

To illustrate this effect, in Figs. \ref{fig:poincare}(a,b) we select four distinct locations \{A,B,C,D\}, corresponding to modes with $k_m/k = 1.2 +  \{0,0.5,1,1.5\}i$, varying the amount of losses. The polarization ellipse for the electric field of the evanescent tail of such a mode is plotted in Fig. \ref{fig:poincare}(c), noting that the local polarization of the electric field tilts and depends strongly on the losses of the waveguide.

\section{Near-field coupling dipole optimization}

\begin{figure*}[!htbp] 
    \includegraphics[width=1\textwidth]{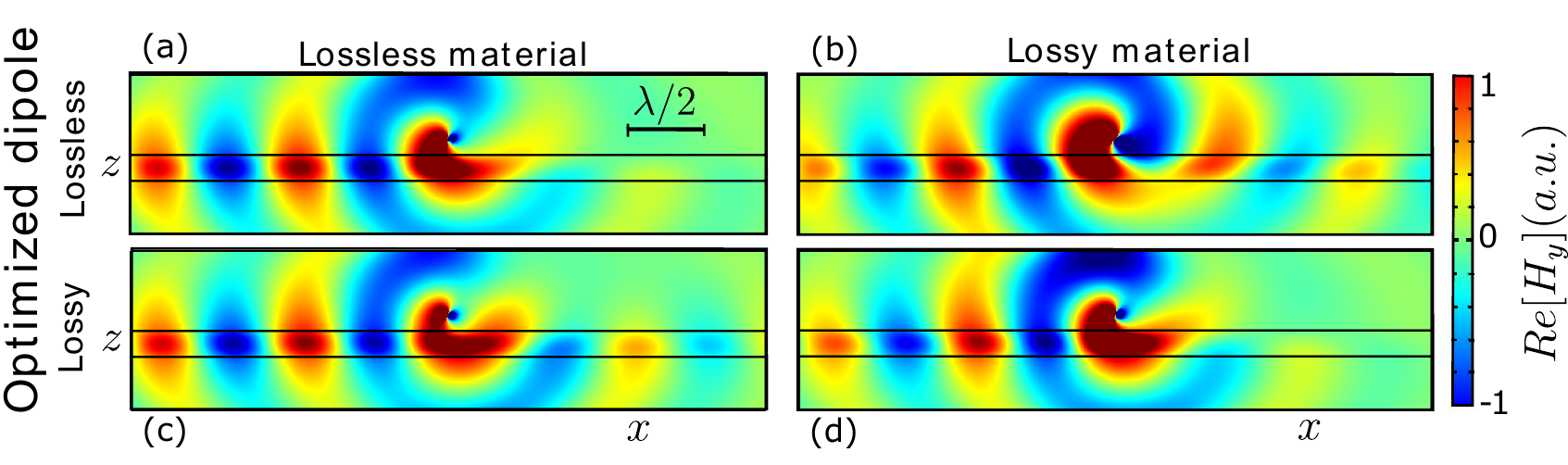}
    \caption{Dielectric slab waveguide with thickness $t=\lambda/6$ placed on free space, excited by a dipole source placed at a distance $d=\lambda/10$ from the dielectric, with a varying dipole optimization (up/bottom) and amount of dielectric losses (left/right). In the left column, a lossless material is considered, with refractive index $n_1=2$, while in the right column a lossy material with $n_2=2+0.3i$ is used. The dipoles were optimized via Fermi's golden rule for near-field selective vectorial coupling to the left side. The upper row has a dipole $\mathbf{p}_1 \approx (1.11,0,0.49i) \propto \mathbf{k}^*_{\mathrm{lossless}} $ while the lower row has $\mathbf{p}_2 \approx (1.10,0,0.11+0.47i) \propto \mathbf{k}^*_{\mathrm{lossy}}$. Clear unidirectionality is shown in (a) and (d) with theoretically contrast ratio 1:0 whereas in (b) and (c) an ``undesired'' back-excitation is observed, with an expected contrast of 1:0.14. Color scale (arbitrary units) is the same for all plots. For the simulation, COMSOL Wave Optics module was used.}
    \label{fig:comsol}
\end{figure*}

Now, the fact that the polarization of the evanescent electric field in the waveguide depends on losses has evident implications for unidirectional coupling of dipoles near the surface. Consider a point-like dipole with polarization given by $\mathbf{p}$=$(p_{x},p_{y},p_{z})$ placed in the evanescent region of the slab waveguide at a height $d$ above the dielectric slab waveguide of thickness $t$. As is known \cite{pakscience}, suitably optimized elliptical dipoles can be used to achieve unidirectional excitation of the guided modes. Here we ask how this is affected by the losses in the waveguide.

Fig. \ref{fig:comsol} shows a numerical simulation of the effect of losses on dipolar directional excitation. Fig. \ref{fig:comsol}(a) corresponds to the known case of a lossless waveguide being excited by an optimized elliptical dipole, exhibiting perfect directionality. Fig. \ref{fig:comsol}(b) shows exactly the same dipole, but with losses added to the waveguide. One can see that, due to the presence of losses, the dipole is no longer perfectly unidirectional, i.e., it does not exhibit a 100\% contrast ratio between left and right excitation. This is because the polarization of the evanescent tails of the waveguides have changed, and hence the optimization of the dipole must take this into account. In Fig. \ref{fig:comsol}(d), the dipole polarization has been adjusted to account for the losses, and this time one sees, indeed, a perfect directionality. This dipole, whose polarization is adjusted for losses, will not work in the lossless waveguide as shown in Fig. \ref{fig:comsol}(c). The results convincingly show that dipole directionality must necessarily take losses into account if one wants to achieve perfect directionality. Next we explain how to deduce this, following two complementary methods: Fermi's golden rule, and dipole angular spectrum.

In the context of dipolar coupling to waveguides, the Fermi's golden rule approach states \cite{fermi,fermi2} that the excitation amplitude of a mode with electric field $\mathbf{E(r)}$ by a dipole $\mathbf{p}$ located at $\mathbf{r}_0$ is proportional to $\mathbf{p}^* \cdot \mathbf{E}(\mathbf{r}_0)$, such that the intensity is proportional to $\abs{\mathbf{p}^* \cdot \mathbf{E}(\mathbf{r}_0)}^2$. Following this approach, one can achieve perfect contrast directionality of excitation simply by choosing a dipole polarization that cannot couple to the evanescent wave propagating along one direction in the waveguide. This is achieved when $\mathbf{p} \propto \mathbf{k}^*$; with this condition, following Gauss' law requirement from Eq. \ref{eq:generalreq}, we can see that $\mathbf{p}^* \cdot \mathbf{E}_\mathrm{ev} \propto \mathbf{k}\cdot\mathbf{E}_\mathrm{ev} = 0$. In our simplified 2D case, this means $\mathbf{p} \propto (k_x^*,0,k_z^*) = (\pm k_m,0,\sqrt{k^2-k_m^2})^*$ with the plus or minus sign determining which of the two directions of mode propagation, right or left, we wish the dipole to not couple to. The polarization of the optimized dipole therefore depends directly on the losses of the waveguide, via $k_m$, and this is the optimized dipole used in the numerical simulations of Figs. \ref{fig:comsol}(a,d).

The above argument using Fermi's golden rule fully explains why taking losses into account is important for dipole directionality, but next we will also analyze the same phenomenon from the dipole angular spectrum approach. This is often used as an alternative explanation to directionality - giving the same results, but offering a different perspective, as it reveals that directionality can be a property of the dipole itself, independently of the waveguide mode \cite{pakscience,originspin}. The directionality of the dipole can be then associated with a zero amplitude at a specific location in the angular spectrum of the dipole source. This specific location is determined by the waveguide mode. This approach offers an intuitive way to design the directionality of multimode waveguides \cite{amplit} by setting sources with zero amplitudes at the angular spectrum location corresponding to the propagation constant of each mode. However, this approach comes with a problem when considering lossy modes, because the propagation constant of the mode is a complex number, while the angular spectrum of a source $\mathbf{E}^\mathrm{dipole}(k_x,k_y)$ is defined on the real plane of transverse wavevectors $(k_x,k_y)$, as follows \cite{Novotny2006}:

\begin{equation} \label{eq:angularspectrumdefinition}
\mathbf{E}^\mathrm{dipole}(\mathbf{r}) = \iint_{-\infty}^{\infty} \mathbf{E}^\mathrm{dipole}(k_x,k_y) e^{i(k_x x + k_y y + k_z z)} dk_x dk_y
\end{equation}

\noindent where $\mathbf{E}^\mathrm{dipole}(\mathbf{r})$ are the spatial fields of a dipole in free space, and $k_z = \pm \sqrt{k^2-k_t^2}$, with $k_t^2 = k_x^2 + k_y^2$ being the transverse wavevector and choosing the sign of $k_z$ depending on the one of $z$. 

The question we ask is, can we study the angular spectrum of the dipole beyond real values of $k_x$ and $k_y$ to study its coupling to lossy modes? Can we define and calculate an angular spectrum defined for complex values in the transverse momentum plane? The definition of the angular spectrum, in Eq. \ref{eq:angularspectrumdefinition}, requires an integral in the real plane $(k_x,k_y)$, however, nothing is stopping us from taking the known analytical form of the angular spectrum of a dipole, which is defined in terms of $k_x$ and $k_y$, and calculating it for complex values of $k_x$. The angular spectrum of an electric dipole source with dipole moment $\mathbf{p}$ is well known (see a concise derivation on Appendix B):

\begin{equation} \label{eq:angularspectrumdipole}
\mathbf{E}^\mathrm{dipole}(k_x,k_y) = \frac{i}{8 \pi^2 \varepsilon} \frac{k^2}{k_z^{(+)}} \left[ (\mathbf{p}\cdot \mathbf{\hat{e}}_p)\mathbf{\hat{e}}_p + (\mathbf{p}\cdot \mathbf{\hat{e}}_s)\mathbf{\hat{e}}_s\right]
\end{equation}

\noindent where $k_z^{(+)}$ indicates taking the sign of $k_z$ that corresponds to positive $z$, $\varepsilon$ is the electric permittivity of the medium, and the two unit vectors $\mathbf{\hat{e}}_p = (\frac{k_x k_z}{k k_t}, \frac{k_y k_z}{k k_t}, - \frac{k_t}{k})$ and $\mathbf{\hat{e}}_s = (-k_y/k_t,k_x/k_t,0)$ represent the p-polarized and s-polarized unit vectors \cite{angularmomen,epes}, remembering to choose the sign of $k_z$ according to whether we are calculating the field in upper hemisphere with $z>0$ or the lower one, where $z<0$.

\begin{figure}[ht]
 
    \centering
    \includegraphics[width=8.6cm]{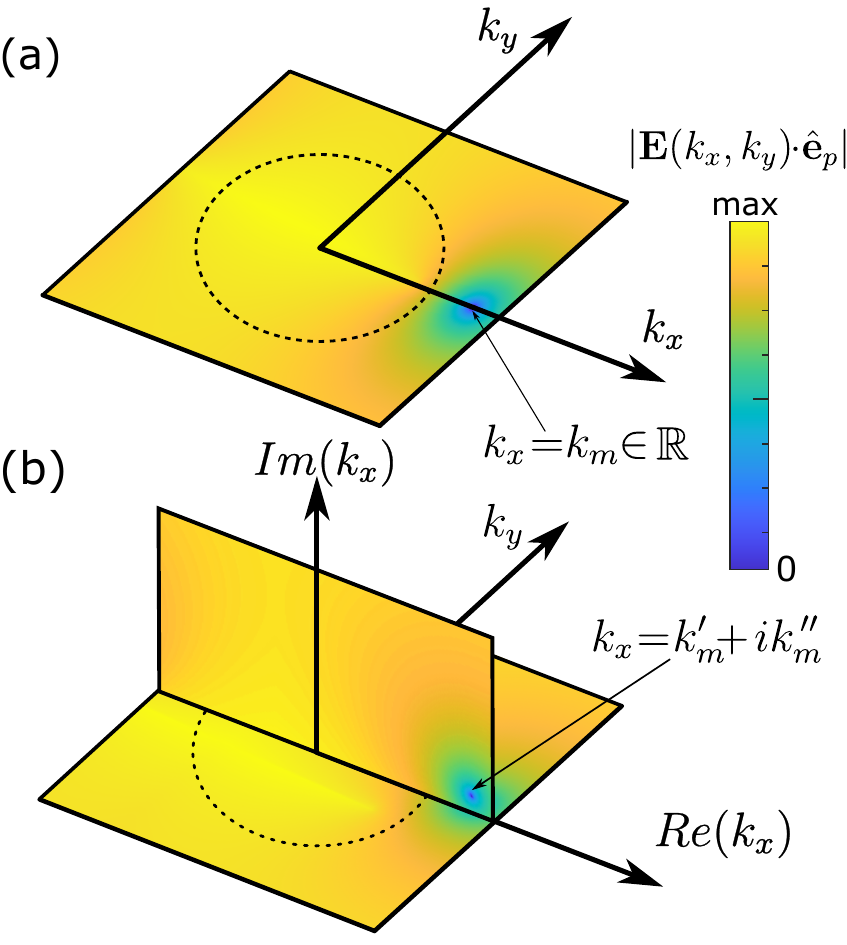}
\caption{Angular spectrum $|\mathbf{E}(k_x,k_y)\cdot\mathbf{\hat{e}}_p|$ for a dipole optimized to show perfect contrast directionality for a p-polarized mode that is (a) lossless with $k_m \in \mathbb{R}$ and (b) lossy  with $k_m = k_m' + i k_m''$ (so $k_m \in \mathbb{C}$). In (b) the angular spectrum does not show zero amplitude at any location on the $(k_x,k_y)$ real plane (the traditional domain of the angular spectrum), but it does show a zero amplitude if we calculate the spectrum in the complex $k_x$ plane.}
\label{fig:manolo}
\end{figure}

If one designs a dipole $\mathbf{p}$ optimized for generating a perfect directionality inside a lossless waveguide mode with a certain real value of propagation constant $k_m$, such as the dipole in Fig. \ref{fig:comsol}(a), then the dipole angular spectrum shows a zero amplitude at $(k_x,k_y) = (k_m,0)$, as shown in Fig. \ref{fig:manolo}(a). This was known since the early designs of directional dipoles \cite{pakscience} but in that same work, the presence of losses was identified as a challenge for unidirectionality. The broadening of the waveguide mode's spatial Fourier spectrum in the real $k_x$ axis due to the losses suggested that one cannot design a dipole source to achieve perfect directionality in a lossy waveguide. After all, where should we place this zero in order to cover the entire broadened range of wavevectors spanned by the lossy mode? The answer is that we need to reinterpret the broadening of the mode as a shift of its position away from the real axis and going into the complex domain of $k_x$. Then it is still possible to design a dipole whose angular spectrum has a zero on the exact location of the mode within the complex domain, enabling the design of perfect directionality even for lossy waveguides, as shown in Fig. \ref{fig:manolo}(b). The dipole designed using this method matches exactly with the one designed using Fermi's golden rule previously. In this figure, the angular spectrum is plotted, using Eq. \ref{eq:angularspectrumdipole}, for a dipole that is designed, as described above, to show optimized directionality on a lossy waveguide. Instead of calculating the spectrum on the real $(k_x,k_y)$ plane, as is conventional for angular spectra, we have also calculated the spectrum in the Argand plane of $k_x$. Interestingly, while the angular spectrum does not show any zero amplitude in the real plane of $(k_x,k_y)$, it does show a zero amplitude at the complex point $k_x = k_m' + i k_m''$, as clearly seen in the figure. This indicates that, although the procedure to calculate the fields from the angular spectrum (Eq. \ref{eq:angularspectrumdefinition}) involves using only the real values of $k_x$ and $k_y$, the information contained in the angular spectrum $\mathbf{E}(k_x,k_y)$ is still meaningful when one considers complex values for the arguments $k_x$ and $k_y$, at least in terms of predicting the source's unidirectionality.

\section{Discussion}

Using an analogy with the phenomenon of spin-momentum locking, in which the spin and polarization of the evanescent wave depends on the propagation direction, we see that the spin and polarization also depends on the losses of the waveguide, showing a phenomenon that we could call polarization-loss locking. It is very interesting to note that the losses of a waveguide can be analytically derived from simply looking at the polarization of the mode at a single fixed point (and vice versa). This allows one to deduce the amplitude decay or spatial gradients of a mode simply with a local precise measurement of polarization, not even requiring the measurement of polarization over a small neighborhood. The local polarization at every point is uniquely mapped, with a one-to-one correspondence, to the complex propagation constant.

We also showed how this polarization-loss relation must be taken into account when designing near-field dipole directionality and carefully described how this is consistent with existing frameworks of dipole directionality. In particular, it required us to stretch the definition of angular spectra, intriguingly showing some evidence that the calculation of angular spectra in complex variables might have physical significance. 

We note that the mode polarization is especially sensitive when the propagation constant is near the threshold with propagating waves $k_m = k_m^0 \gtrapprox 1$, corresponding to weakly confined modes. In that case, tiny changes in either the real part $(k_m = k_m^0 +\epsilon)$ or the imaginary part $(k_m = k_m^0 + i \epsilon)$ (with $0<\epsilon \in \mathbb{R}$) of the propagation constant will result in comparably large changes in the corresponding polarization of the evanescent wave, which can be translated into significant variations in dipolar coupling to waveguided modes. This suggests that a modulation of intensity could be achieved using directional dipole sources near a waveguide whose real or imaginary part of refractive index are changed via a material non-linearity, as well as a potential way to encode information on spatial variations of optical losses of a material, whose readout can be realized optically via directional sources.

It is always interesting to see physical phenomena arise when extending variables that are typically considered real into the complex domain. Further research is envisaged exploiting the unidirectional coupling not only in lossy waveguides, as done here, but also lossy surrounding media as well, when the waveguide is embedded in a complex refractive index, or other possibilities such as near-zero index materials \cite{nzi}. In this case, the wave-equation condition $\mathbf{k}\cdot\mathbf{k} = k^2$ becomes even more interesting because the right-hand-side wavenumber may become complex itself, so $\mathbf{k}$ acquires more degrees of freedom, and new exotic nanophotonic phenomena could arise.

\section*{Acknowledgment}
S. P.-P. wants to thank I. Cajiao-Valle for the computer- assisted graphic design tutorials, J. J. Kingsley-Smith  for his advice on computational plot representation and fruitful  discussion, D. Mart\i'nez-Rubio for mathematical discussion in the Poincar\'e sphere, Dr. M. F. Picardi for the suggestions on the dielectric unidirectional model and Dr. L. Wei for the useful advice given in the numerical simulation of the materials. This work was supported by European Research Council Starting Grant ERC-2016-STG-714151-PSINFONI and EPSRC (UK). The authors declare no competing financial interest.

\bibliography{output}{}

\appendix

\section{Stokes parameters dependence on the propagation constant}

In order to calculate the Stokes parameters of the evanescent tail electric field employed to plot the paths on the Poincar\'e sphere in Fig. \ref{fig:poincare}, we used the definition of Stokes parameters \cite{perrin1942,estockes} but applied in the $(x,z)$ plane where the electric field lies:

\begin{equation} \label{eq:stokes}
    \begin{pmatrix}S_0 \\ S_1 \\ S_2 \\ S_3 \end{pmatrix}=
    \begin{pmatrix}
    E_x\cdot E^*_x+E_z\cdot E^*_z\\
   E_x\cdot E^*_x-E_z\cdot E^*_z\\
    E_x\cdot E^*_z+E_z\cdot E^*_x\\
   iE_x\cdot E^*_z-iE_z\cdot E^*_x\\
    \end{pmatrix}.
\end{equation}

According to the conditions in Eq. \ref{eq:simplifiedreq}, it is easy to see that, once $k_m$ is fixed, and hence $k_x = k_m$, then we can calculate $k_z = \sqrt{k^2-k_x^2}$ using the Helmoltz condition (such that $\mathrm{Im}(k_z)\geq 0$) and then the electric field is restricted by the transversality condition to $\mathbf{E}_0 = A_0(\mathbf{\hat{y}}\times\mathbf{k}) = A_0 (-k_z,0,k_x)$ where $A_0$ is an arbitrary scaling factor. Substituting this into the Stokes parameters in Eq. \ref{eq:stokes} results in parametric paths of polarization along the Poincar\'e sphere as a function of mode propagation constant $\mathbf{S}(k_m) = (S_0,S_1,S_2,S_3)$ used to generate the paths in Fig. \ref{fig:poincare}:
\begin{equation} \label{eq:stokesparametric}
    \begin{pmatrix}S_0 \\ S_1 \\ S_2 \\ S_3 \end{pmatrix}=\begin{pmatrix}
    |k_z|^2 + |k_x|^2\\
  |k_z|^2 - |k_x|^2\\
    -2 \mathrm{Re}\left(k_x^* k_z\right)\\
   \phantom{-}2 \mathrm{Im}\left(k_x^* k_z\right)\\
    \end{pmatrix}=|k_m|^2   
    \begin{pmatrix}
    |\eta|^2+1\\
    |\eta|^2-1\\
    -2 \mathrm{Re}\left(\eta\right)\\
    \phantom{-}2 \mathrm{Im}\left(\eta\right)\\
    \end{pmatrix},
\end{equation}

\noindent where superscript $*$ means complex conjugation, $|\cdot|$ means complex euclidean modulus, and we have used the ratio \cite{blio}:

\begin{equation}
\eta(k_m)=\frac{k_z}{k_x}=\frac{\sqrt{k^2-(k'_m + i k''_m)^2}}{k'_m + i k''_m},
\end{equation}

\noindent with special care taken to always take the square root sign that guarantees a positive imaginary part of $k_z$, to ensure a physically meaningful evanescent wave above the waveguide (the opposite sign should be used for the evanescent wave below the waveguide). From here, many analytical limits in the Poincar\'e sphere can be mathematically obtained. In the propagating plane wave case $k_m/k=1$, $\eta$ becomes null, so $S_2/S_0=S_3/S_0=0$ and analogously $S_1/S_0=-1$. In the limit of growing lossless mode propagation constant $k'_m \to \infty$, one can calculate that $\lim_{k'_m \to \infty}\eta = i$. The same limit can be found for the case of unbounded mode losses $\lim_{k''_m \to \infty}\eta = i$. In both cases, therefore, the polarization tends to $S_1/S_0=S_2/S_0\to0$ and $S_3/S_0\to1$, corresponding to the upper pole of the Poincar\'e sphere as depicted in Fig. \ref{fig:poincare} from the main text.

\section{Angular spectrum of an electric dipole}
The angular spectrum of a dipole source in a homogeneous medium has been derived previously in the literature \cite{Mandel1995,Novotny2006,angularmomen,epes,tespic} but here we present a very concise derivation. Our starting point is the vector potential of a dipole \cite{Novotny2006,japan} given as $\mathbf{A} = -i \omega \mu \mathbf{p} \frac{e^{i k r}}{4 \pi r}$ where $\omega$ is the angular frequency, $\mu$ is the magnetic permeability, $k = n \omega / c$ is the wavenumber and $r=|\mathbf{r-r_0}|$ is the radial distance to the dipole position $\mathbf{r}_0$, which we take as $\mathbf{r}_0=\mathbf{0}$ below. The electromagnetic fields can be calculated from the potential as $\mathbf{H} = \frac{1}{\mu} \nabla \times \mathbf{A}$ and $\mathbf{E}=-\frac{1}{i \omega \varepsilon} \nabla \times \mathbf{H}$. Combining these well known definitions we can write the electric field of a dipole as:

\begin{equation} \label{eq:Erdipole}
    \mathbf{E(r)} = \nabla \times \nabla \times \left( \frac{1}{\varepsilon} \mathbf{p} \frac{e^{i k r}}{4 \pi r}\right).
\end{equation}

In order to find the angular spectrum, we need to perform a double spatial Fourier integral. To do this, we can use the well-known Weyl identity \cite{Mandel1995}:

\begin{equation} \label{eq:weyl}
    \frac{e^{i k r}}{r} = \frac{i}{2\pi}\iint_{-\infty}^{\infty} \frac{e^{i \mathbf{k} \cdot \mathbf{r}}}{k_z^{(+)}} d k_x d k_y,
\end{equation}

\noindent where $\mathbf{r}=(x,y,z)$, $\mathbf{k}=(k_x,k_y,k_z)$, with the sign of $k_z = \pm \sqrt{k^2-k_t^2}$ chosen depending on whether we are taking $z>0$ or $z<0$ such that $\mathrm{Im}(k_z) > 0$ or $\mathrm{Im}(k_z) < 0$ respectively, and where $k_z^{(+)}$ specifies having to take the sign for $z>0$. Substituting Eq. \ref{eq:weyl} into \ref{eq:Erdipole}, we can evaluate the curl $(\nabla \times)$ operators inside the integral, which due to the harmonic dependence become $(i\mathbf{k} \times)$ operators, resulting in:

\begin{equation} \label{eq:Edipolespectrumstep1}
    \mathbf{E(r)} = \iint_{-\infty}^{\infty} \frac{1}{\varepsilon}\frac{i}{8 \pi^2 k_z^{(+)}}  \left[ i\mathbf{k} \times i\mathbf{k} \times \mathbf{p} \right] e^{i \mathbf{k} \cdot \mathbf{r}} d k_x d k_y.
\end{equation}

By simple comparison between Eq. \ref{eq:Edipolespectrumstep1} and the definition of the angular spectrum in Eq. \ref{eq:angularspectrumdefinition} of the main text, we immediately identify the angular spectrum of the dipole as the terms multiplying the exponential. This finalises our derivation of the angular spectrum of the dipole. However, one may note that the angular spectrum is a vector quantity $\mathbf{E}(k_x,k_y)$, hence in practice it is useful to decompose it into its components in some basis.

A useful basis is the spherical basis, aligned with the relative orientation of the wavevector, defined by the unit vectors $\{\mathbf{\hat{e}}_k,\mathbf{\hat{e}}_p,\mathbf{\hat{e}}_s\}$ where we define $\mathbf{\hat{e}}_k = \mathbf{k}/k$, $\mathbf{\hat{e}}_s = (\mathbf{\hat{z}} \times \mathbf{k})/\sqrt{(\mathbf{\hat{z}} \times \mathbf{k})\cdot(\mathbf{\hat{z}} \times \mathbf{k})}$ and $\mathbf{\hat{e}}_p = \mathbf{\hat{e}}_s \times \mathbf{\hat{e}}_k$, such that one can check they form an orthonormal basis $\mathbf{\hat{e}}_i \cdot \mathbf{\hat{e}}_j = \delta_{ij}$, with the interesting feature that the basis vectors are in general complex-valued and yet our definition of orthonormality does not involve complex conjugation. This is possible thanks to the wave-equation, Eq. \ref{eq:generalreq} in the main text, which results in $\mathbf{\hat{e}}_k\cdot\mathbf{\hat{e}}_k=1$. Thanks to this orthonormality, any vector can be expressed in terms of its components $\mathbf{p} = (\mathbf{p}\cdot\mathbf{\hat{e}}_k)\mathbf{\hat{e}}_k + (\mathbf{p}\cdot\mathbf{\hat{e}}_s)\mathbf{\hat{e}}_s + (\mathbf{p}\cdot\mathbf{\hat{e}}_p)\mathbf{\hat{e}}_p $. Therefore, making use of the fact that $\hat{\mathbf{e}}_k \times \hat{\mathbf{e}}_s = -\hat{\mathbf{e}}_p$ and $\hat{\mathbf{e}}_k \times \hat{\mathbf{e}}_p = \hat{\mathbf{e}}_s$ or using Lagrange's formula for triple product, it is straightforward to show that $i\mathbf{k} \times i\mathbf{k} \times \mathbf{p} = -k^2 (\mathbf{\hat{e}}_k \times \mathbf{\hat{e}}_k \times \mathbf{p}) = k^2\left[ (\mathbf{p}\cdot\mathbf{\hat{e}}_s)\mathbf{\hat{e}}_s + (\mathbf{p}\cdot\mathbf{\hat{e}}_p)\mathbf{\hat{e}}_p \right]$ is simply a projection of $\mathbf{p}$ into the space orthogonal to $\mathbf{\hat{e}}_k$. Substituting this into Eq. \ref{eq:Edipolespectrumstep1} we arrive directly at:

\begin{equation} \label{eq:Edipolespectrumstep2}
    \mathbf{E(r)} = \iint_{-\infty}^{\infty} \frac{i k^2}{8 \pi^2 \varepsilon k_z^{(+)}} \mathbf{p} \left[ (\cdot \mathbf{\hat{e}}_p)\mathbf{\hat{e}}_p + (\cdot \mathbf{\hat{e}}_s) \mathbf{\hat{e}}_s\right] e^{i \mathbf{k} \cdot \mathbf{r}} d k_x d k_y,
\end{equation}

\noindent which completes the derivation of Eq. \ref{eq:angularspectrumdipole} in the main text.
\end{document}